\documentclass[12pt]{revtex4}
\usepackage{graphicx}

\begin{document}

\title{Non-Stationary Correlation Matrices And Noise}

\author{Andr\'e C. R. Martins}
\affiliation{GRIFE -- Escola de Artes, Ci\^encias e Humanidades USP}
%Av. Arlindo B\'etio, 1000 CEP 03828-000
%S\~ao Paulo - SP - Brazil\\
\email{amartins@usp.br}

\begin{abstract}
The exact meaning of the noise spectrum of eigenvalues of the correlation matrix is discussed. In order to better understand the possible phenomena behind the observed noise, the spectrum of eigenvalues of the correlation matrix is studied under a model where most of the true eigenvalues are zero and the parameters are non-stationary. The results are compared with real observation of Brazilian assets, suggesting that, although the non-stationarity seems to be an important aspect of the problem, partially explaining some of the eigenvalues as well as part of the kurtosis of the assets, it can not, by itself, provide all the corrections needed to make the proposed model fit the data perfectly.


\end{abstract}

\maketitle

\section{Introduction}

Given a set of data, determining what consists of a signal, relevant to the problem one is trying to solve, is a very important problem. What part of the data is actually noise can depend on the question but it is also determined by the model one uses to fit the data. In the analysis of financial series, determining what part of the data, if any, is just noise and what part involves a real signal can, in principle, make an important difference in the decisions to be taken. In particular, estimating covariance (or correlation) matrices is an important part of traditional methods of portfolio choice~\cite{markowitz}.

In the last years, this problem has been addressed by using the methods of Random Matrix Theory (RMT)~\cite{wigner,mehta}. The first results showed that most of the eigenvalues of the correlation matrix (except for a few large ones) had basically the same behavior as the eigenvalues of a random matrix~\cite{laloux99,plerou99,plerou2002}. Several independent tests, for different markets confirmed those findings~\cite{utsugi2004,maslov}, even though, recently, there as been found some evidence that the fit is not perfect and that the noise eigenvalues seem to be a little larger than they should be~\cite{kwapien2006}. A possible proposed explanation for this behavior is that it might be caused by small pair wise correlations between all the assets~\cite{malevergne}.  It has also been observed that, despite the fact that they are usually considered as non-existent and only noise, correlations can be measured in the random part of the eigenvalue spectrum~\cite{burda,burdafinancial}.

However, despite the fact that, apparently, much of the correlation matrix seems to be caused by noise, the effects of this noise on actual portfolio optimization might not be as important than it would appear at first, apparently causing only small deviations from optimal asset choices~\cite{pafka}. Also, RMT seems to provide a good way to deal with the problem of the high dimensionality involved in most financial applications~\cite{pafka2004a,pafka2004b,bouchaud2005}.

For small windows of time, it has been verified that different behaviors of the eigenvalues can be observed at different points of time~\cite{drosdz2000, drosdz2001}. In particular, decreases usually happen together with a strong collective eigenstate, while this collective phenomenon is not so important during increases. This suggests that non-stationarity might be an important aspect of the eigenvalue problem, as it is in general for financial time series~\cite{bouchaudpotters}, despite the fact that most models, for simplicity, do not include it. It has been shown however, that some of the stylized facts of financial time series, such as the long range dependence in volatility and integrated GARCH~\cite{starica2004} and the changing covariance structure~\cite{starica2006} can actually be explained as effects of the non-stationarity of the series.

In this article, the problem of non-stationarity and its consequences for the eigenvalues of the correlation matrix will be investigated. The question of the meaning of the noise, measured by the eigenvalues of the correlation, will be discussed and a toy model for the non-stationarity of the parameters that shows that some, but not all, of the eigenvalues can be caused by this non-stationarity will be presented. This model, for a suitable choice of the parameters have many zero eigenvalues in the stationary case, making it clearer what effect the non-stationarity has on the eigenvalues.

\section{Noise and eigenvalues}

Suppose the real covariance matrix, from where the time series of the prices of $N$ assets are drawn, is given by the $N\times N$ matrix $\mathbf\Sigma$. In order to estimate it, one has $T$ observations of returns $G_i(t)$, where $i=1,\cdots,N$, made at different points in time $t$, from where the sample covariance matrix $\mathbf C$ can be obtained by
\begin{equation}
C_{ij}=\frac{1}{T}\sum_{k=1}^{T}G_i(k)G_j(k) - \frac{\sum_{k=1}^{T}G_i(k)}{T}\frac{\sum_{k=1}^{T}G_j(k)}{T},
\end{equation}
where the last term, the multiplication of the average observed returns for assets $i$ and $j$ ($\bar{G}_i=\frac{\sum_{k=1}^{T}G_i(k)}{T}$ and similarly for $j$) can only be dropped if we have renormalized the returns so that they have zero average. From $\mathbf C$, the correlation matrix $\mathbf R$ can be easily obtained by $R_{ij}=C_{ij}/\sqrt{C_{ii}C_{jj}}$.

In most applications of RMT in the analysis of the financial series, the properties of $\mathbf R$ are studied, with particular emphasis to the distribution of the eigenvalues $\lambda_i$ of $\mathbf R$. Basically, except for a few large eigenvalues, most of the observed values of $\lambda_i$ can be fit reasonably well by the eigenvalues of a random matrix.

The observed property that the eigenvalues of a random matrix seem to fit reasonably well most of the observed eigenvalues in real market realizations of the prices is usually interpreted as meaning that those eigenvalues are most likely due to noise. On the other hand, correlations are measurable in this ``noise'' region of the spectrum~\cite{burda,burdafinancial} and it makes sense to ask if, although small, the bulk of the eigenvalues has some meaning other than noise. 

It should also be noted that, in a strict mathematical sense, this interpretation of the smaller eigenvalues as noise, is not correct, if by noise one means that there are only a few non-zero eigenvectors, while all other eigenvectors of the covariance matrix are null. Suppose that all correlations between market prices are actually caused by a finite number $e$ of eigenvectors, while all other $N-e$ eigenvalues are exactly zero. Under these circumstances, random realizations of this stochastic process will not generate observations with noisy eigenvalues. The real $e$ eigenvalues can be estimated with increasing precision as $T$ grows, but, even for small values of $T$, the remaining eigenvalues will actually be observed as exactly zero. This is due to the fact that the covariance matrix is singular and assigns zero probability to any observations that would correspond to everything not covered by the existing eigenvalues. As a matter of fact, even generating those realizations can be a problem, if not dealt with care, since, as $|{\mathbf \Sigma}|=0$, the traditional multivariate normal density does not exist~\cite{anderson}, if one insist on working with $N$ variables. That is, the problem is only well defined if one uses the eigenstate structure to generate the realization and, by doing so, no new eigenvalues appear.

This means that no noise appears in the traditional sense, unless the noise is actually generating new eigenvectors; however, when a covariance matrix is generated by RMT, non-zero eigenvalues do appear. This means that the meaning of those eigenvalues must be better understood. When generating the real covariance matrix by RMT, there is no causal reason for the eigenvalues and this fact is not in dispute. But the generated $\mathbf \Sigma$ matrices do have non-zero eigenvalues. Despite the lack of causal apparent reasons, those eigenvalues correspond to actual correlations, as indicated by the results that show that they are actually measurable. RMT actually provides the average of many possible covariance matrices, each of them with a full spectrum. That means that, unlike the common use of the term noise, one can not, in principle, treat the non random eigenvalues as the only real ones and ignore all the others.
If the interpretation of them as noise is to be fully recovered, the origin of this noise must be better understood. In this paper, the possibility that the non-stationarity of the parameters might be causing the noise is investigated.

\section{The Model}

In order to better understand if the noise eigenvalues can be generated by some simple random mechanism, a model with zero eigenvalues is needed. And, since it has been observed that non-stationarity can play an important part in this problem, the model should allow the parameters to change, while providing a realistic covariance matrix, that is, one that obeys all the requirements for such a matrix.  

At a given instant, $t$, the return and covariance of the $N$ assets will be modeled by an $N\times M$ matrix $\mathbf \Phi$, with components $\varphi_{ij}$, where $i=1,\cdots , N$ represents the different assets and where each value of $j$, $j=1,\cdots , M$, $M\geq3$, can be seen as a different possible state of the system. That is, $\mathbf \Phi$ can be seen as a collection of $M$ typical vectors $\mathbf \varphi$, with $N$ components. Given $\mathbf \Phi$, the average return vector $\mathbf \mu$, the covariance matrix $\mathbf \Sigma$ and the correlation matrix $\mathbf P$ will be given by

\begin{equation}\label{parametrization}
\mu_{i} = E\left[ \varphi_{i} \right]= \frac{1}{M} \sum_{j=1}^{M} \varphi_{ij}
\]
\[
\Sigma_{il} = \frac{1}{M} \sum_{j=1}^{M} \varphi_{ij}\varphi_{lj} - \mu_{i}\mu_{l},
\]
\[
P_{il}=\frac{\Sigma_{il}}{\sqrt{\Sigma_{ii}\Sigma_{ll}}}
\end{equation}
\noindent
and the observed returns $r_{t}$, at instant $t$, are generated, as usual, by a multivariate normal $N(\mu,\Sigma)$ likelihood. It should be noted here that the components of $\mathbf \Phi$ are the real parameters of the model, not the average return or the covariance matrix.

Therefore, the model will have $N\times M$ parameters, used to obtain the $\frac{N(N+1)}{2}$ values of the covariance matrix $\Sigma$ plus the $N$ average returns $\mu_i$ (even if $\mu_i$ are chosen to be zero, $N$ parameters will have to be chosen so that the average returns are actually zero). The conflict between a model with fewer parameters and, therefore, easier to estimate, and one that describes better all the observed facts should be at the heart of the choice of an appropriate value for $M$ for a real application of the model. But, since we are interested in understanding the noise by using this model first, a choice of a small value of $M$, that will generate few non-zero eigenvalues, is the most appropriate here.

Notice that, as long as $M$ is reasonably smaller than $N$, there will be less parameters than in the usual average-covariance parametrization. That means that the components in the covariance matrix thus obtained are not all independent. As a consequence, several eigenvalues might be actually exactly equal to zero. More exactly, the model will have $M-1$ non-zero eigenvalues and $N-M+1$ zero eigenvalues. The apparently missing eigenvalue ($M-1$ eigenvalues, instead of $M$) is due to the fact that one of the vectors $\mathbf \varphi$ is used to specify the model average. Even if it is taken to be zero, one vector is needed to ensure it is actually zero, leaving $M-1$ vectors available to determine the eigenvalues and eigenvectors of the correlation matrix. This property of the model is actually useful for this investigation. Since we are interested in studying the effect of non-stationarity, those effects can be more easily noticed when the comparison is made against a case where the stationary eigenvalues are exactly zero.

An interesting feature of this model is that one can easily impose a temporal dynamics on the $\varphi$ parameters. By doing so, the parameters for the normal distribution of the returns will be altered but all the characteristics of the covariance matrix will be automatically respected, since the covariance is actually calculated in Equation~\ref{parametrization} as the covariance of the parameters. Also, as the parameters are basically typical observations, it is not difficult to interpret them.

In order to implement the non-stationarity, the model can be altered by making the parameters $\varphi$ change with time, following a Markov process, where $E[{\mathbf \varphi}_{t+1}]={\mathbf \varphi}_{t}$ (the index $i$ referring to the assets was omitted by writing $\varphi$ as the vector $\mathbf \varphi$). 

\begin{equation}\label{eq:nonstationary}
{\mathbf \varphi}_{t+1}={\mathbf \varphi}_{t}+{\mathbf\epsilon},
\end{equation}
\noindent
where each component of the ${\mathbf\epsilon}$ vector follows a $N(0,\sigma_\epsilon^{2})$distribution. Notice that this temporal dynamics doesn't preserve the average returns, but that is not a problem for our current analysis, since we are interested only in the correlation matrix. For an average preserving dynamics, if a value $\epsilon$ is added to a component $\varphi_{ij}$, it must be subtracted from another component $\varphi_{ij'}$, so that $\sum_j \varphi_{ij}$ will remain the same. Also, a common source of confusion in this non-stationarity is that the $\varphi_{ij}$ are not observations, but true parameters of the model. Since they will have new values randomly drawn at each instant, the parameters are now a function of time, $\varphi_{ij}(t)$, and, hence, the non-stationarity of the model. More than that, since the correlations are calculated from them, the correlation structure will change with time and, therefore, introduce new eigenvectors. That is, $\sigma_\epsilon$ is not simply a scale parameter, as it will cause the whole correlation strusture to change with time. 

If we were actually interested in the parameters $\varphi_{ij}(t)$, it is true that $\sigma_\epsilon$ will just associate a normal distribution to them, with $\sigma_\epsilon$ as scale. But notice that, even if the initial $\varphi_{ij}(0)$ were known for sure, Equation~\ref{eq:nonstationary} will make it unknown for the next time period. That is, the real parameters of the model change with time and an estimation of the problem requires estimating them at each point in time. More than that, the dynamics on the parameters will cause the correlation structure to change with time, meaning the process will not be covariance-stationary~\cite{hamilton}.

We should notice that, since the correlation matrix is calculated from the parameters as if the parameters were a sample of observed results, the properties of a correlation matrix are preserved by construction.

\section{Simulation Results}

\begin{figure}
	\centering
		\includegraphics[width=0.75\textwidth]{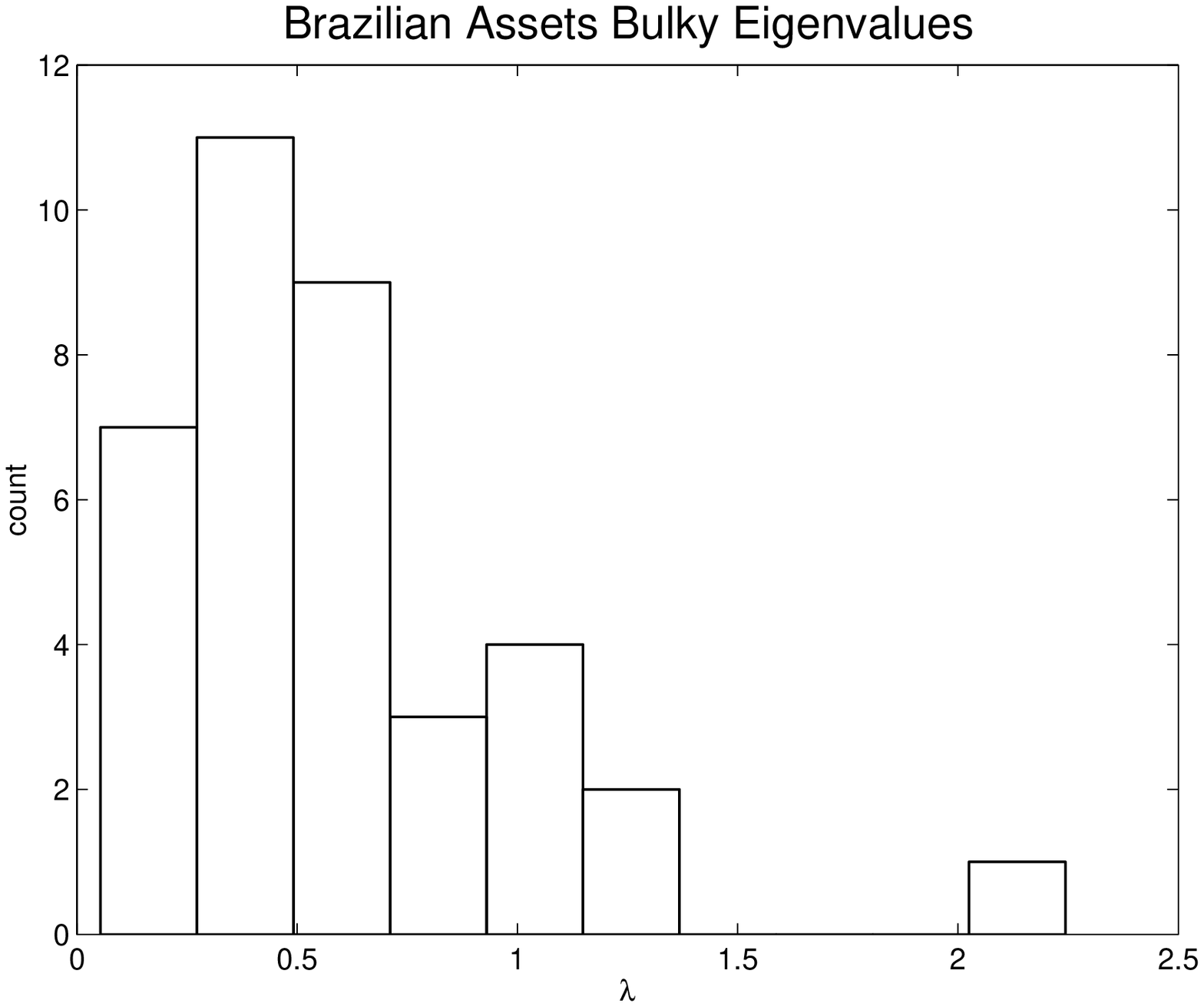}
	\caption{Bulky eigenvalues of $N=38$ Brazilian stocks, calculated from the correlation matrix. The main eigenvalue (16.2) is not shown in the graph.}
	\label{fig:braseigen}
\end{figure}

Several simulations were run in order to test the model. In order to have some real data to compare with, the returns of $N=38$ Brazilian stocks were observed daily from January, 5th, 2004 to July, 28th, 2006, for a total of $T=644$ observations. The correlation matrix obtained from this set of data shows one large eigenvalue (16.2) and 37 other bulk eigenvalues as shown in Figure~\ref{fig:braseigen}. The average (over the assets) observed kurtosis is 1.93, with standard deviation of 3.12.

Simulations of 644 daily returns with $N=38$ assets, $M=2$ (one non-zero eigenvalue) and values of $\sigma_\epsilon$ ranging from 0 to 0.45 (where 1 would mean a random walk with a daily standard deviation of 1\% for each of the values of the parameters) were performed and each case run 100 times in order to obtain average results. The initial values of the parameters were drawn randomly from a multivariate normal distribution with zero average and covariance matrix equal to the sample covariance matrix of the Brazilian assets, so that the comparison between the simulated results and the real data made sense. 

\begin{figure}
	\centering
		\includegraphics[width=0.80\textwidth]{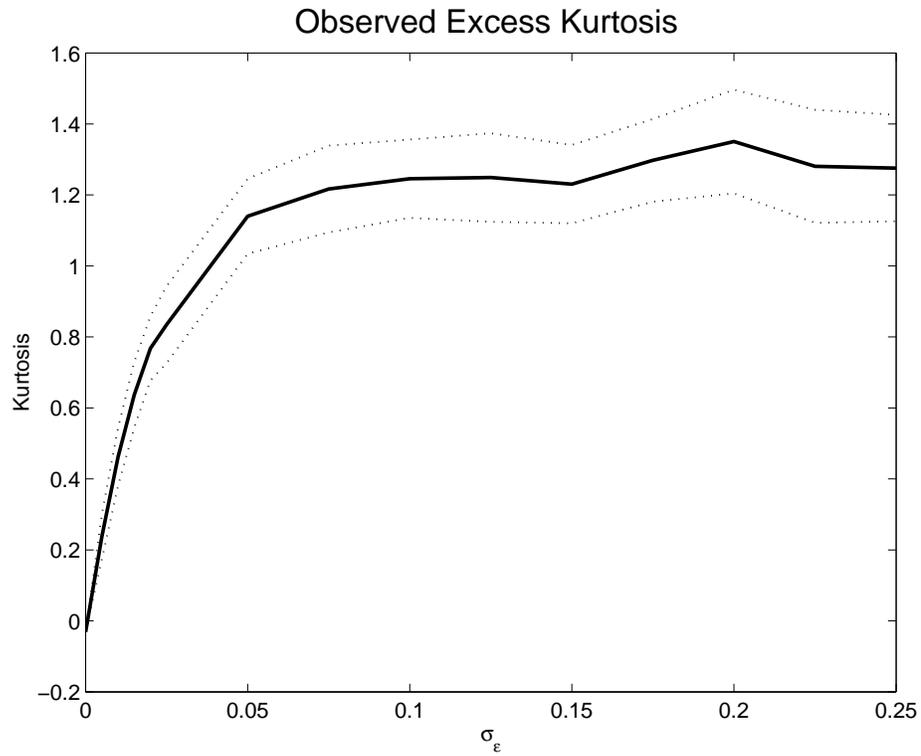}
		\caption{Average observed kurtosis of assets over 100 simulations as a function of the random walk standard deviation $\sigma_\epsilon$ (the dotted lines show one simulated standard deviation from the average simulated value).}
	\label{fig:kurtosis}
\end{figure}

The observed kurtosis, as a function of $\sigma_\epsilon$ can be seen in Figure~\ref{fig:kurtosis}. As soon as some non-stationarity is introduced in the problem ($\sigma_\epsilon\neq 0$), the kurtosis becomes greater than zero, and it stabilizes around 1.2 as soon as $\sigma_\epsilon$ is between 0.05 and 0.1. Simulations for larger values of $\sigma_\epsilon$, as well as fr a larger number of assets, $N=200$, were performed, showing this stabilization. Notice that, although the kurtosis is not as large as that observed in Brazilian stocks, even a small non-stationarity cause the appearance of fat tails. In order to understand the scale of the non-stationarity, $\sigma_\epsilon=0.05$ means that, after 100 days, the expected value of each $\varphi_{ij}(t)$ would have drifted with a standard deviation of $0.05\sqrt{100}=0.5$, causing the expected return to drift with a standard deviation of $0.5/\sqrt{M}\approx 0.35$, a significant drift.

\begin{figure}
	\centering
		\includegraphics[width=0.80\textwidth]{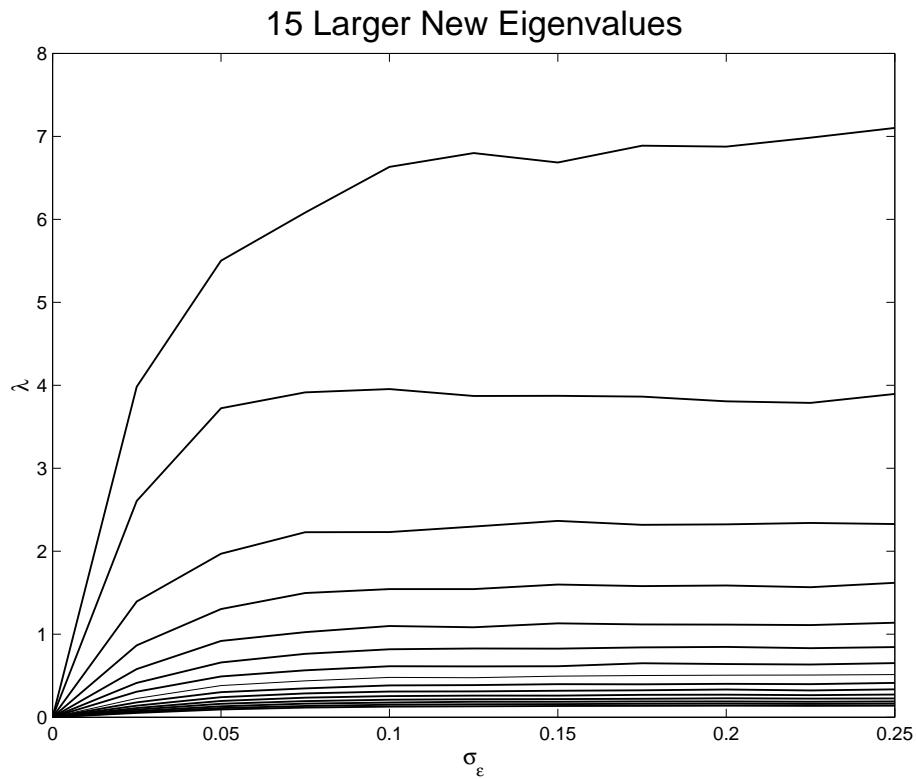}
	\caption{Simulated eigenvalues generated from non-stationarity as a function of the random walk standard deviation $\sigma_\epsilon$. The graph shows the 15 larger new values, the remaining 22 smaller ones (close to zero) are not shown to avoid a mess of lines in the $\lambda=0$ region.}
	\label{fig:eigen}
\end{figure}

The eigenvalues of the observed returns were also recorded and the average values for those eigenvalues calculated. In order to obtain the averages, the eigenvalues were ranked from larger to smaller, so that the averages refer to the largest eigenvalue, second largest and so on. The 15 larger new eigenvalues can be seen, as a function of $\sigma_\epsilon$, in Figure~\ref{fig:eigen}. These are the main eigenvalues generated due to the non-stationarity of the model and were zero in the stationary case (as can be seen as $\sigma_\epsilon\rightarrow 0$). It is interesting to notice that a few bulky eigenvalues are generated even for smaller values of $\sigma_\epsilon$ and that, as $\sigma_\epsilon$ gets larger, the eigenvalues exhibit the same stable behavior as the kurtosis. The largest bulky Brazilian eigenvalue was around 2, as can be seen in Figure~\ref{fig:braseigen} and such a value is obtained as a non-stationary eigenvalue for $\sigma_\epsilon$ as small as $\sigma_\epsilon=0.01$ (for this value, the observed average kurtosis was 0.46). However, as $\sigma_\epsilon$ becomes larger, Figure~\ref{fig:eigen} shows that the largest new eigenvalue becomes too big, between 6 and 7 for $\sigma_\epsilon$ values above 0.1. This suggests that the non-stationarity can explain the appearance of the main bulky eigenvalue, since it does not need to be strong in order to generate it. Repeating the analysis of the last paragraph, $\sigma_\epsilon=0.01$ means a drift for the returns, after 100 days of about 0.07 (returns were measured as percentage).

\begin{figure}
	\centering
		\includegraphics[width=0.80\textwidth]{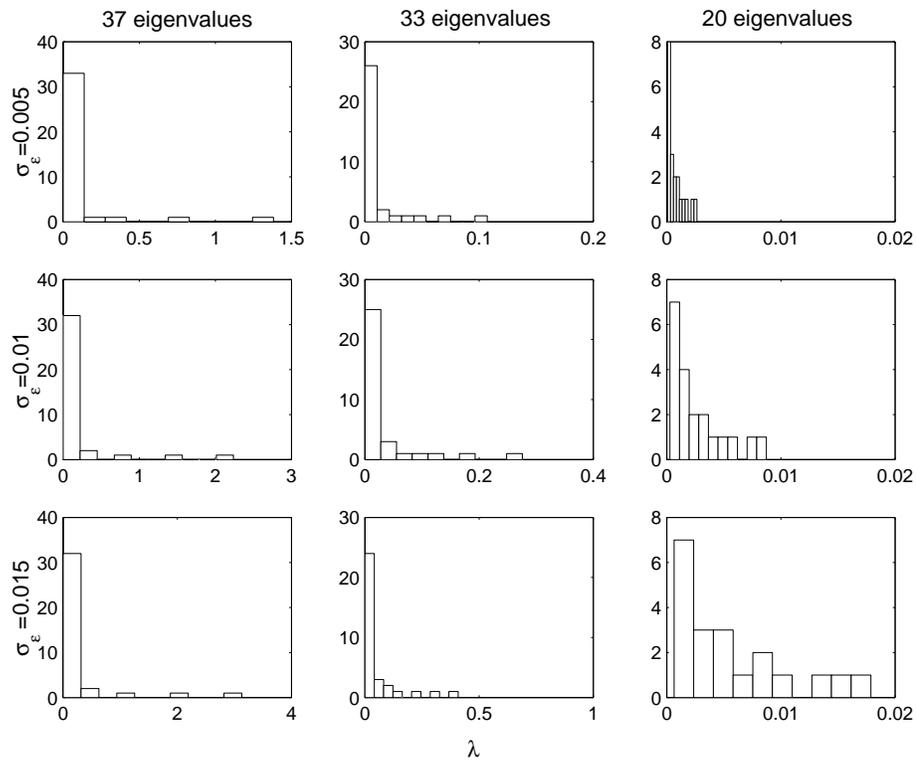}
	\caption{Histogram of the simulated new eigenvalues (37, 33 and 20 smallest non-stationary eigenvalues) generated from non-stationarity, for different values of the random walk standard deviation $\sigma_\epsilon$.}
	\label{fig:n38hist}
\end{figure}

Figure~\ref{fig:n38hist} shows the histogram of the new eigenvalues for different values of $\sigma_\epsilon$, 0.005, 0.01, and 0.015. Notice that most eigenvalues are actually quite small, as shown in Figure~\ref{fig:eigen}, but, as the non-stationarity becomes larger, the histogram moves away from zero (the smallest observed eigenvalues were, respectively, $7.5\cdot10^{-5}$, $2.7\cdot10^{-4}$, and $6.7\cdot10^{-4}$). It is interesting to compare those values with the Mar\u{c}enko-Pastur result~\cite{marcenkopastur}. For $N=38$ and $T=644$, the eigenvalues predicted by RMT lie between 0.57 and 1.54, while the eigenvalues observed due to non-stationarity are not confined to that region and, in fact, most of them are observed bellow the 0.57 value. This means that, although the model proposed here can account for some of the eigenvalues, it can not explain them all.

One interesting consequence of these observations is that, since zero eigenvalues would mean zero eigenvalues in the sample covariance matrix and since non-stationarity can only explain part of the eigenvalue spectrum, this seems to support the idea that there are correlations even in the smaller eigenvalues. Those correlations might not be large and, in part, the eigenvalues can be due to non-stationarity. But at least part of them seems to be real.
On the other hand, the fact that all corrections introduced by the non-stationarity are in the right direction, seems to show that the change of the parameters in time is something that should be taken into account in every description of this problem and further inquiry into consequences of the non-stationarity is needed.

\bibliography{Nonstationarity}

\end{document}